# Alleviating the trade-off between coincidence time resolution and sensitivity using scalable TOF-DOI detectors

Yuya Onishi and Ryosuke Ota*

Central Research Laboratory, Hamamatsu Photonics K. K., Hamamatsu, Japan

These authors contributed equally to this work.

* Author to whom any correspondence should be addressed.

**Abstract**

Coincidence time resolution (CTR) in time-of-flight positron emission tomography (TOF-PET) has significantly improved with advancements in scintillators, photodetectors, and readout electronics. Achieving a CTR of 100 ps remains challenging due to the need for sufficiently thick scintillators—typically 20 mm—to ensure adequate sensitivity because the photon transit time spread within these thick scintillators impedes achieving 100 ps CTR. Therefore, thinner scintillators are preferable for CTR better than 100 ps. To address the trade-off between TOF capability and sensitivity, we propose a readout scheme of PET detectors. The proposed scheme utilizes two orthogonally stacked one-dimensional PET detectors, enabling the thickness of the scintillators to be reduced to approximately 13 mm without compromising sensitivity. This is achieved by stacking the detectors along the depth-of-interaction (DOI) axis of a PET scanner. We refer to this design as the cross-stacked detector, or xDetector. Furthermore, the xDetector inherently provides DOI information using the same readout scheme. Experimental evaluations demonstrated that the xDetector achieved a CTR of 175 ps full width at half maximum (FWHM) and an energy resolution of 11% FWHM at 511 keV with $3 \times 3 \times 12.8$ mm$^3$ lutetium oxyorthosilicate crystals, each coupled one-to-one with silicon photomultipliers. In terms of xy-spatial resolution, the xDetector exhibited an asymmetric resolution due to its readout scheme: one resolution was defined by the 3.2 mm readout pitch, while the other was calculated using the center-of-gravity method. The xDetector effectively resolves the trade-off between TOF capability and sensitivity while offering scalability and DOI capability. By integrating state-of-the-art scintillators, photodetectors, and readout electronics with the xDetector scheme, achieving a CTR of 100 ps FWHM alongside high DOI resolution becomes a practical possibility.

Keywords: positron emission tomography, time-of-flight, coincidence time resolution, depth-of-interaction

## 1. Introduction

Positron emission tomography (PET) utilizes the coincidence detection of two back-to-back 511 keV gamma rays to estimate the annihilation position along a line of response (LOR) between scintillation crystals within the system detector ring. Its imaging performance can be significantly enhanced by incorporating time-of-flight (TOF) information from coincident events (Schaart 2021). TOF information spatially constrains the annihilation position on an event-by-event basis, improving the signal-to-noise ratio (SNR) of the PET image compared with a non-TOF PET system, as described by the following equation:

$$SNR_{\mathrm{Gain}} = \frac{SNR_{\mathrm{TOF}}}{SNR_{\mathrm{nonTOF}}} = \sqrt{\frac{2 \cdot D}{c \cdot \Delta t}}, \quad (1)$$

where $D$ and $c$ represent the subject diameter and the speed of light, respectively, and $\Delta t$ denotes the coincidence time resolution (CTR), a representative metric for TOF information. Currently, the CTR of commercial PET scanners is approximately 200 ps full width at half maximum (FWHM) (Van Sluis *et al* 2019, Zhang *et al* 2024), which limits the spatial uncertainty to approximately 3 cm along the LORs. This translates into an SNR gain of 3.7-fold for a subject of diameter 40 cm.

A long-standing milestone for TOF-PET is achieving a system with a CTR of 100 ps FWHM, which could provide a 5-fold improvement in SNR. This advancement has the potential to revolutionize not only reconstructed image quality but also lesion detectability, quantitativeness, and radiation dose reduction in clinical practice (Schaart 2021). The timing performance of a scintillation-based PET detector is determined by the detection chain of a gamma-ray (Lecoq 2017). The detection chain primarily comprises factors such as interaction position, scintillation kinetics, photon detection probability, single-photon time resolution (SPTR) of the photodetector, electronic jitter, and photon transit time spread (PTTS) in the scintillation crystal. Recent technological advancements in scintillators, photodetectors, and readout electronics have enabled CTRs of 100 ps FWHM or less in lutetium-based scintillators (Nadig *et al* 2023, Mariscal-Castilla *et al* 2024) using short crystals. A typical clinical PET scanner requires a crystal with at least 20 mm length for adequate detection efficiency of 511 keV gamma rays (Hsu *et al* 2017, Van Sluis *et al* 2019). However, targeting a CTR of 100 ps FWHM in a detector configuration with sufficient length, the influence of variations in PTTS is not negligible as the PTTS is equivalent to tens of picoseconds (Gundacker *et al* 2014, Cates *et al* 2015). Thus, the trade-off between TOF capability and sensitivity renders the detectors complicated. Recently, a strategy of side readout (Cates and Levin 2018, Gonzalez-Montoro *et al* 2021, Pourashraf *et al* 2021, Cates and Levin 2023) has been considered to achieve a better CTR while maintaining high sensitivity using 20 mm-thick scintillators. However, increased number of photodetectors, which increase the total cost of a system, and reduced packing fraction pose another challenge.

As a different approach, we propose a readout scheme that retains the typical scintillator-photodetector configuration to reduce the trade-off between TOF capability and sensitivity. The core idea is to orthogonally stack two one-dimensional PET detectors, with one longitudinal side uncovered by a reflector and facing each other's bare side. This design enables the reduction of the scintillator length without compromising sensitivity by stacking the detectors along the depth-of-interaction (DOI) axis of a PET scanner. We refer to this design as the cross-stacked detector, or xDetector. Additionally, the xDetector provides scalable DOI information while keeping the same readout scheme, resulting in a high-performance TOF-DOI PET detector that meets the stringent demands of the PET research field (Yoshida and Yamaya 2024). DOI information enhances spatial resolution, which becomes more effective for long axial field-of-view scanners that have seen widespread adoption in recent years (Vandenberghe *et al* 2020). Therefore, the xDetector also has the potential to enable state-of-the-art PET scanners with enhanced spatiotemporal resolution.

## 2. Concept of xDetector

The concept and structure of the proposed xDetector are summarized in figure 1. A rectangular scintillator is coupled to a silicon photomultiplier (SiPM) or multi-pixel photon counter (MPPC) (Ota 2021). Four sides of the scintillator are covered by reflectors, while one longitudinal side remains bare; the bottom is coupled to an SiPM, as shown at the top of figure 1(a). Multiple detectors, such as the three single detectors shown in figure 1, are aligned and grouped along the short-axis direction to form a detector group. The two detector groups are then orthogonally stacked, with the bare sides facing each other via air coupling, as depicted in the bottom of figure 1(a).

The expected behavior of scintillation photons when a gamma-ray event enters the detector unit is visualized in figure 1(b). When a gamma ray interacts with channel 1, the scintillation photons ideally do not propagate to the adjacent channel 2 due to the reflector but can propagate through the air coupling to the detector in the upper layer (top view of figure 1(b)). For a lutetium orthosilicate (LSO) crystal with a refractive index of approximately 1.8, total reflection easily occurs due to the refractive index difference at the air interface. Consequently, most scintillation photons are detected by channel 1, while some are detected by channel 5 (side view of figure 1(b)). If the interaction occurs at the center of channel 1, an equal amount of light ideally leaks out to channels 4 and 6. Based on the absolute and relative pulse heights of the acquired signal waveforms (figure 1(c)), it can be considered that a photoelectric event occurred in channel 1 and below channel 5.

Figures 1(d) and 1(e) illustrate the potential shape of a module comprising 12 detector units and a PET scanner ring, respectively. When the SiPM readout pitch is set to 4.2 mm, the scintillator length can be reduced to 12.6 mm. Despite this reduction, a 1-inch module can still be formed by rotating and combining the detector units. As the scintillator length used in conventional PET scanners is typically 20 mm, the reduced length of 12.6 mm is expected to improve timing resolution (Gundacker *et al* 2014). By stacking the detector units and defining the z-axis (DOI axis) as shown in figure 1(d), the DOI resolution will be 4.2 mm, and the sensitivity will remain comparable to that of a conventional PET detector despite the shorter scintillator length. Notably, the detector performance is independent of the number of layers stacked, as proposed by Peng *et al* (2019). Thus, users can select an arbitrary number depending on their applications.

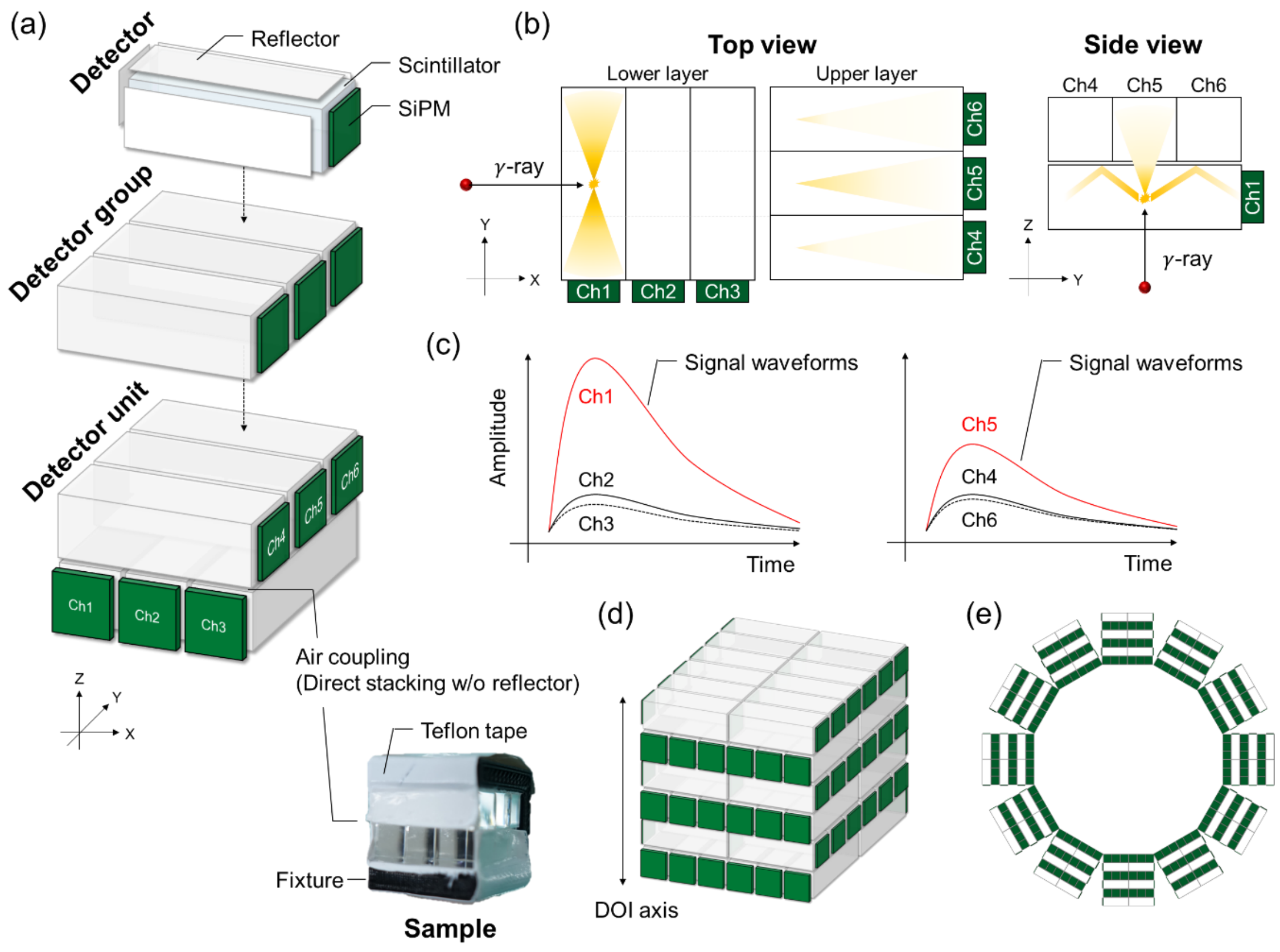

**Figure 1.** Schematic illustration of the proposed xDetector. (a) Design and sample of the xDetector, comprising orthogonally stacked in two detector groups via air coupling. The sample shown is prior to SiPM attachment. (b) Expected behavior of scintillation photon propagation and (c) signal waveforms when a gamma-ray event enters the detector unit of the channel 1. (d) Potential shape of a module comprising 12 detector units and (e) a system detector ring for a PET scanner.

## 3. Experimental setup and performance measurements

We developed two prototypes of xDetectors with varying sizes. The first group comprises four

LSO crystals (3 × 3 × 12.8 $mm^3$; EPIC crystal, China) coupled to an MPPC (S14161-3050HS; Hamamatsu Photonics K.K., Japan), while the second group includes three LSO crystals (4 × 4 × 12.6 $mm^3$; EPIC crystal) coupled to an MPPC (S14161-4050; Hamamatsu Photonics K.K.). Each LSO crystal has its four sides wrapped with an enhanced specular reflector (ESR) and a single layer of Teflon tape. The two detector groups are orthogonally stacked with their bare sides facing each other. To avoid any decrease in reflectivity caused by adhesives such as optical grease, custom fixtures were created using a 3D printer (Raise3D Technologies Inc, USA), as shown in the sample in figure 1(a). To assess the performance of the xDetectors, we measured the CTR, linearity, energy resolution, and longitudinal spatial resolution. The performance other than the longitudinal spatial resolution was compared with that of conventional single detectors of the same length (12.8 or 12.6 mm) and a length of 20 mm, each wrapped with ESR and Teflon tape on five sides.

### 3.1. Coincidence time resolution (CTR)

Figure 2 shows the experimental setup used to measure the CTR of the xDetectors and conventional single detectors. A point source of $^{22}Na$, which emits back-to-back 511 keV gamma rays, was used. The source was placed at a sufficient distance from the xDetectors to ensure that the gamma rays uniformly impinge along the longitudinal side of the upper scintillator, as the xDetectors are designed to be irradiated uniformly along this side, as illustrated in figure 1(e). In this study, the upper detector output closest to the point source was defined as channel 1.

For the coincidence measurement, only a specific channel was readout and fed into an oscilloscope (DSO-404A; Keysight Technologies, USA) with a sampling rate of 20 GS/s and a bandwidth of 4.2 GHz. As a reference detector, a 3 × 3 × 10 $mm^3$ LYSO crystal (EPIC crystal) coupled to an MPPC (S13360-3075; Hamamatsu Photonics K.K.) with high-frequency electronics was used (Ota and Ote 2024). The single timing resolution of the reference detector was 111.2 ± 0.8 ps FWHM. For the timing signals from the reference detector and the specific channel of the xDetector, the vertical ranges of the oscilloscope were limited to precisely monitor the rising edges of the timing signals. On the other hand, energy signals from the two detectors were used to trigger the oscilloscope, but not displayed and recorded to maintain the maximum sampling rate of the oscilloscope. Energy threshold was determined from the pulse height of the energy signal and was approximately set at the valley between the photo peak and the Compton edge.

The coincidence waveforms collected from the oscilloscope were analyzed to determine the optimal timing pick-off thresholds. Detection timing was established by sweeping the threshold level for timing pick-off to obtain the best CTR at overvoltages of 3, 5, 7, and 8 V. The CTR was evaluated by fitting a Gaussian function to the histogram of the time difference between the xDetector and the reference detector, with the FWHM of the Gaussian function representing the

CTR.

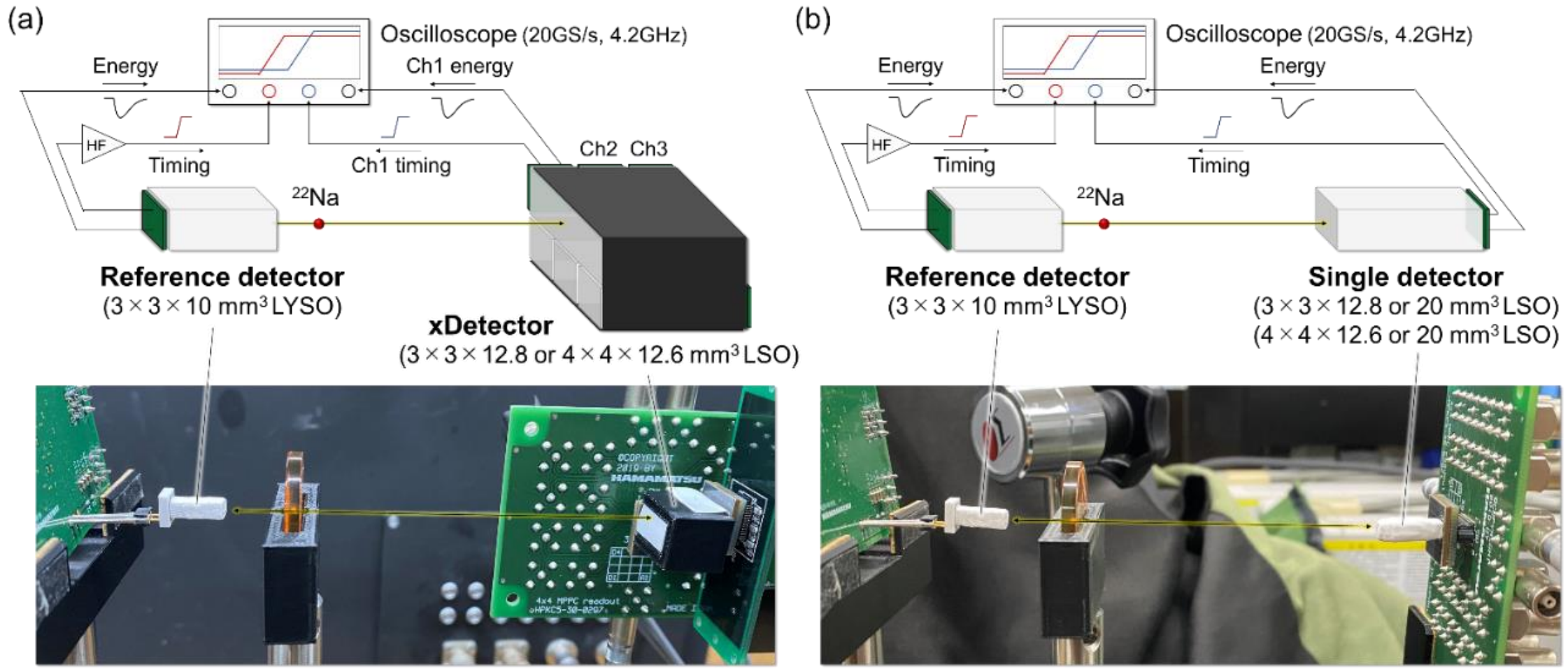

**Figure 2**. Experimental setups for measuring the CTR of (a) xDetectors with channel 1 signal readout and (b) conventional single detectors in which five sides of the crystal is wrapped by an ESR and a Teflon tape. The upper part shows illustrations of the setups, while the lower part presents photographs of the setups. HF refers to high frequency-amplifier readout electronics.

### 3.2. Linearity and energy resolution

MPPCs exhibit a saturation effect that results in a non-linear response to input energy as the number of detected scintillation photons approaches the number of micropixels (Gundacker and Heering 2020). This saturation reduces the MPPC signal, thereby degrading the energy resolution. The relationship between the number of incident photons $N_{\mathrm{photon}}$ and the MPPC output, represented by the number of fired micropixels $N_{\mathrm{fired}}$, is expressed as follows:

$$N_{\mathrm{fired}} = N_{\mathrm{total}}\left\{1 - \exp\left(-\frac{N_{\mathrm{photon}} \cdot PDE}{N_{\mathrm{total}}}\right)\right\}, \tag{2}$$

where $N_{\mathrm{total}}$ denotes the total number of micropixels, and $PDE$ denotes the photon detection efficiency. As the number of incident photons is proportional to the energy of a gamma ray during the scintillation process, the linearity function $f(E)$ is defined as:

$$f(E) = \mathrm{A}\{1 - \exp(-\mathrm{B} \cdot E)\}, \tag{3}$$

where $\mathrm{A}$ and $\mathrm{B}$ denote parameters of the function, and $E$ denotes the energy (Ota *et al* 2017). We can calculate the linearity $l(E)$ at a certain energy using the following equation:

$$l(E) = \frac{f(E)}{\frac{df}{dE}|_{E=0} \cdot E} \cdot 100. \tag{4}$$

The denominator of equation (4) represents the ideal response line, expressed as $\mathrm{AB} \cdot E$. To measure the energy resolution of the detector, the inverse function of equation (3) was employed to translate the voltage signal into energy. A Gaussian + exponential + constant function (where the exponential and constant account for contamination from the Compton edge to the photopeak and the Compton continuum from 1275 keV) was fitted to the translated energy spectra to obtain energy resolutions at 511 keV.

The experimental setup for measuring linearity and energy resolution is almost identical to the CTR measurement shown in figure 2, with only single data collected from the xDetector used. Linearity functions were derived using two peak values corresponding to 511 and 1275 keV from $^{22}$Na.

### 3.3. Longitudinal spatial resolution

Figure 3 displays the experimental setup for measuring the longitudinal spatial resolution of xDetectors. The $^{22}$Na point source was positioned close to the upper scintillator of the xDetector to limit the interaction position along the longitudinal axis by setting up a coincidence measurement with the reference detector placed at a sufficient distance from the source. Interaction positions along the longitudinal direction were controlled by simultaneously stepping both the source and the reference detector using stages with 10 um precision. In conventional scintillation detectors used in PET scanners, the longitudinal axis corresponds to the DOI axis, whereas for the proposed detector, this relationship differs. Notably, the longitudinal spatial resolution of the xDetector corresponds to the spatial resolution along the x- or y-axis, and the DOI resolution is determined by the size of the scintillator, which is 3 or 4 mm in this experiment.

The xDetector was supplied with an overvoltage of 3 and 7 V. MPPC cathode signals were fed into constant fraction discriminators (TC454; Tennelec, USA), and coincidence trigger signals were generated using a coincidence unit (N017; Hoshin Electronics, Japan). The coincidence waveforms from channels 1 to 6 were digitized using a VME module (V1742; CAEN, Italy), which has a bandwidth of 500 MHz and a sampling rate of 1 GS/s.

The coincidence waveforms undergo baseline correction and energy selection. The interaction position $Y$ along the longitudinal side is then estimated using the center of gravity calculation, known as Anger logic (Anger 1964), as follows:

$$Y = \sum_{i=4 \text{ or } 5}^{2N} y_i \cdot w_i \; / \sum_{i=4 \text{ or } 5}^{2N} w_i \, , \tag{5}$$

where $y_i$ denotes the center coordinate of the *i*th column of the MPPC array, and $w_i$ denotes the pulse height of the digitized waveform from the *i*th column readout. $N$ represents the number of columns in the SiPM array ($2N$ = 6 or 8). Moreover, we compared the performance with a ratio-based calculation method (Kang *et al* 2015) that uses the signals at both ends according to the following equation: $Y = (w_4 - w_6) \,/\, (w_4 + w_6)$ or $= (w_5 - w_8) \,/\, (w_5 + w_8)$. In the setup shown in figure 3, three outputs from channels 4 to 6 were used for the center-of-gravity calculation, while two outputs from channels 4 and 6 were used for the ratio-based calculation. The longitudinal spatial resolutions were evaluated by fitting a Gaussian function to a histogram of the interaction positions and were defined as the FWHM of the function.

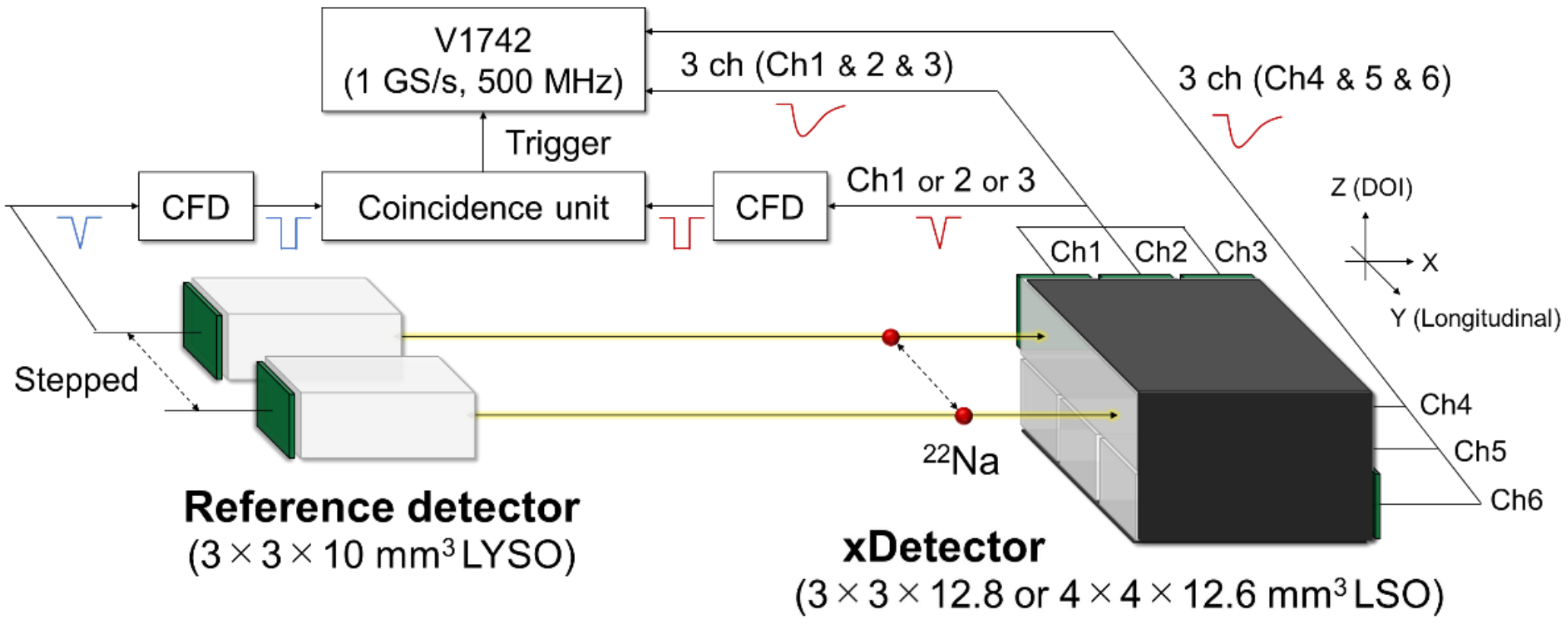

**Figure 3.** Experimental setups for measuring the longitudinal spatial resolution of xDetectors. The positron source and reference detector are simultaneously moved along the longitudinal axis of the upper scintillator. The longitudinal spatial resolution of the xDetector corresponds to the spatial resolution along the y-axis. It should be noted that the conventional DOI resolution does not correspond to the longitudinal spatial resolution, but is determined by the size of the scintillator.

## 4. Results

### 4.1. Coincidence time resolution (CTR)

Figure 4 presents the CTRs for xDetectors and single detectors with 3 and 4 mm$^2$ at different overvoltages. The best CTRs for the xDetectors with 3 and 4 mm$^2$ were 175.3 ± 1.3 ps FWHM for channel 4 at an overvoltage of 7 V, and 187.4 ± 1.7 ps FWHM for channel 3 at an overvoltage of 5 V, respectively. For both sizes, xDetectors achieved similar CTR performance compared to single detectors of the same length, with minimal variation across channels. Moreover, they significantly outperformed the CTRs of single detectors of length 20 mm.

Notably, the scintillator with dimensions 3 × 3 × 20 $mm^3$ used in this experiment was manufactured approximately three years earlier than the other scintillators, and we found that this difference led to variations in scintillator properties such as decay time and relative light yield, even though they were all manufactured by the same company. Therefore, the CTRs of the single detector with 3 × 3 × 20 $mm^3$ shown in figure 4(a) were the corrected values using the analytic CTR expression based on the following equation (Gundacker *et al* 2020):

$$\mathrm{CTR}_{\mathrm{analytic}} = 3.33 \cdot \sqrt{\frac{\tau_{\mathrm{diff}} \cdot (1.57 \cdot \tau_r + 1.33 \cdot \sigma_{\mathrm{SPTR+PTS}})}{\mathrm{PDE} \cdot \mathrm{LTE} \cdot \mathrm{ILY}_{@\mathrm{Energy}}}}, \quad (6)$$

where $\tau_{\mathrm{diff}}$ and $\tau_r$ represent the effective decay time and rise time of the scintillator, $\sigma_{\mathrm{SPTR+PTS}}$ denotes the convolution of the SPTR with the PTTS of the crystal, $\mathrm{LTE}$ is the light transfer efficiency, and $\mathrm{ILY}$ is the intrinsic light yield. The relative to ILY for the 20 mm scintillator compared to the 12.8 mm scintillator ($\mathrm{ILY}_{20}/\mathrm{ILY}_{12.8}$) was estimated to be 1.45. This estimation utilized the relationship between LTE and scintillator length reported in Cates and Levin (2018). The decay time $\tau$ of the scintillators was measured using a time-correlated single photon counting (TCSPC) setup, as illustrated in figure 5(a). The 20 and 12.8 mm crystals exhibited single exponential decays of 37.1 and 46.6 ns, respectively. Based on these measurements, the CTR ratio between the 20 and 12.8 mm crystals was estimated to be 1.35 (= $\sqrt{1.45 \times 46.6 \,/\, 37.1}$), and this value was used for correction.

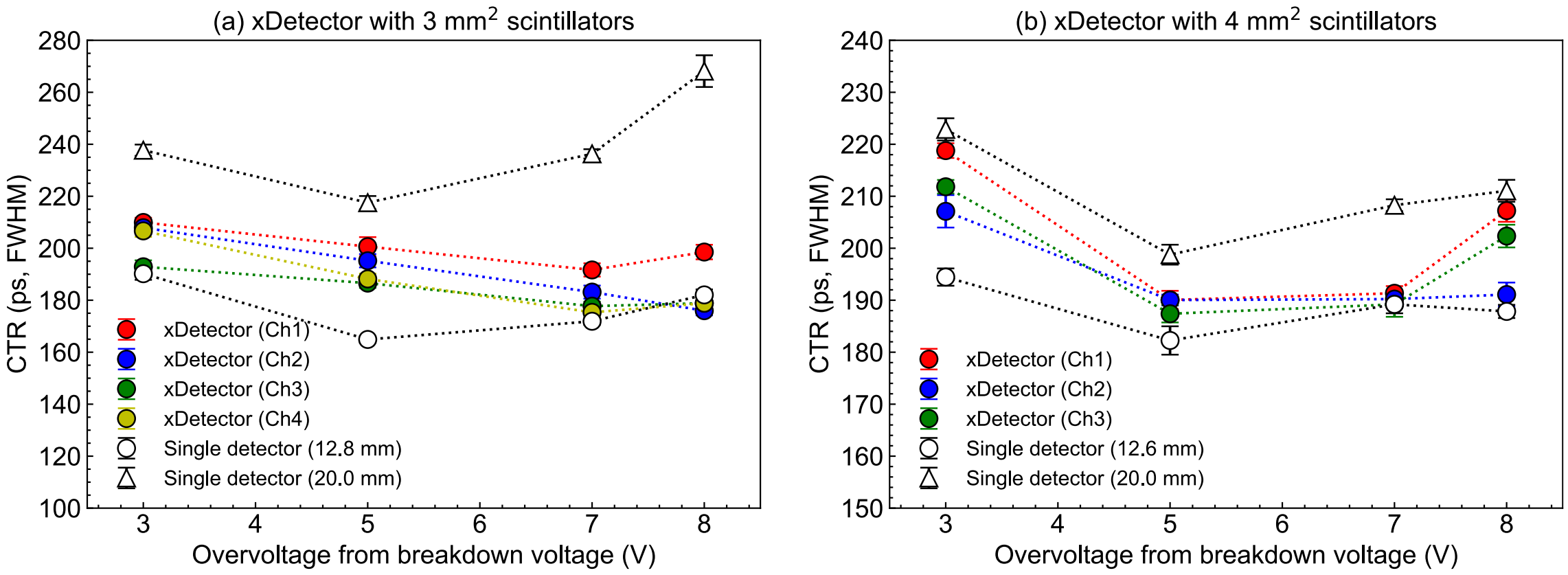

**Figure 4.** CTRs for xDetectors and single detectors with (a) 3 and (b) 4 $mm^2$ at different overvoltages. The best CTRs for the xDetector with 3 and 4 $mm^2$ were 175.3 ± 1.3 ps FWHM for channel 4 at an overvoltage of 7 V and 187.4 ± 1.7 ps FWHM for channel 3 at an overvoltage of 5 V, respectively.

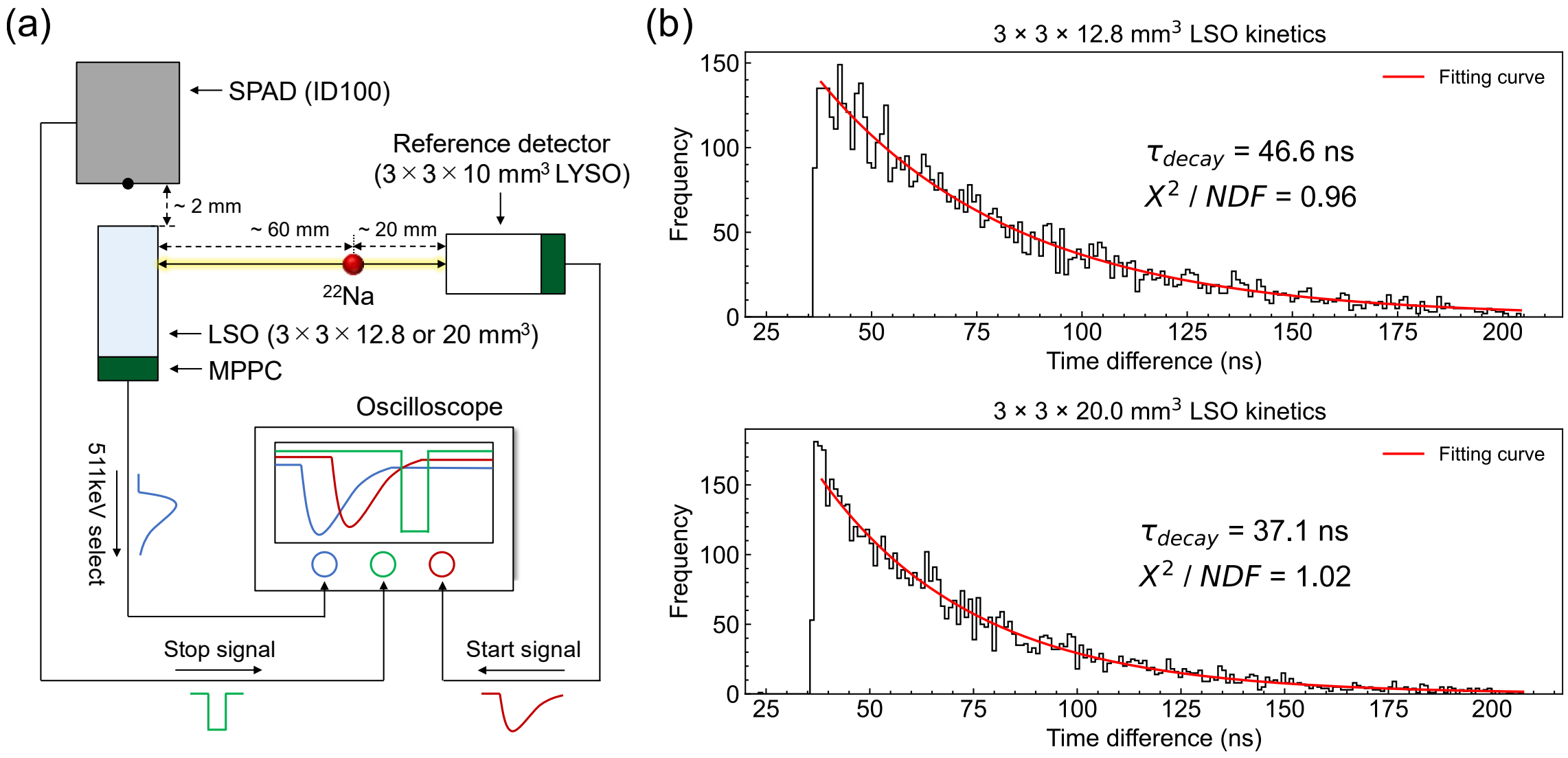

**Figure 5.** (a) Schematic of the TCSPC setup used to measure the intrinsic decay times of LSO crystals of lengths 12.8 and 20 mm. The scintillation kinetics were determined from the distribution of the time differences between the start and stop signals. (b) Scintillation kinetics of LSO crystals of lengths 12.8 and 20 mm, showing single exponential decays of 37.1 and 46.6 ns, respectively. Data collection continued until the fitting performance $\chi^2/NDF$, where $NDF$ is the number of degrees of freedom, approached a value of one.

### 4.2. Linearity and energy resolution

Tables 1 and 2 present the linearity and energy resolution for xDetectors and single detectors with 3 and 4 mm² at different overvoltages. The best linearity at 511 keV for the xDetector with 3 and 4 mm² was 75.96% for channel 1 and 79.38% for channel 3 at an overvoltage of 3 V, respectively, both surpassing conventional single detectors. This improvement is attributed to partially escaped photons interacting with the orthogonally stacked detector group.

The best energy resolution at 511 keV for the xDetectors with 3 and 4 mm² was 11.07% for channel 3 at an overvoltage of 7 V and 9.06% for channel 2 at an overvoltage of 3 V, respectively. Figure 6 further illustrates the linearity curves and energy spectra corrected for non-linearity for channel 2 of the 4 mm² xDetector, which exhibited the best performance at different overvoltages. Increasing the applied voltage was found to degrade both linearity and energy resolution.

**Table 1.** Linearity of xDetectors and single detectors at different overvoltages.

| Crystal size ($mm^3$) | Composition | | Linearity (%) at 511 keV | | | |
|---|---|---|---|---|---|---|
| | | | Vov = 3 (V) | Vov = 5 (V) | Vov = 7 (V) | Vov = 8 (V) |
| 3 × 3 × 12.8 | xDetector | Ch1 | 75.96 | 69.00 | 63.23 | 60.53 |
| | | Ch2 | 73.77 | 69.99 | 61.25 | 59.03 |
| | | Ch3 | 74.71 | 68.12 | 61.93 | 59.97 |
| | | Ch4 | 74.36 | 68.58 | 62.47 | 60.56 |
| 3 × 3 × 12.8 | Single detector | | 68.59 | 61.70 | 55.18 | 54.23 |
| 3 × 3 × 20.0 | Single detector | | 68.80 | 62.37 | 56.30 | 54.79 |
| 4 × 4 × 12.6 | xDetector | Ch1 | 78.81 | 75.17 | 71.04 | 68.81 |
| | | Ch2 | **79.38** | 75.71 | 71.94 | 69.58 |
| | | Ch3 | 79.26 | 74.98 | 71.03 | 68.91 |
| 4 × 4 × 12.6 | Single detector | | 74.46 | 69.94 | 65.60 | 62.71 |
| 4 × 4 × 20.0 | Single detector | | 76.80 | 72.62 | 68.10 | 65.62 |

**Table 2.** Energy resolution of xDetectors and single detectors at different overvoltages.

| Crystal size (mm$^3$) | Composition | | Energy resolution (%) at 511 keV | | | |
|---|---|---|---|---|---|---|
| | | | Vov = 3 (V) | Vov = 5 (V) | Vov = 7 (V) | Vov = 8 (V) |
| 3 × 3 × 12.8 | xDetector | Ch1 | 12.02 | 12.97 | 11.62 | 12.63 |
| | | Ch2 | 12.03 | 11.66 | 11.96 | 11.34 |
| | | Ch3 | 11.87 | 12.40 | 11.07 | 12.46 |
| | | Ch4 | 13.25 | 11.75 | 11.59 | 11.24 |
| 3 × 3 × 12.8 | Single detector | | 11.86 | 11.22 | 11.31 | 13.39 |
| 3 × 3 × 20.0 | Single detector | | 10.87 | 9.35 | 10.00 | 12.33 |
| 4 × 4 × 12.6 | xDetector | Ch1 | 10.22 | 10.74 | 10.81 | 10.67 |
| | | Ch2 | **9.06** | 9.83 | 10.63 | 10.93 |
| | | Ch3 | 9.58 | 9.85 | 10.46 | 10.37 |
| 4 × 4 × 12.6 | Single detector | | 10.93 | 9.80 | 10.06 | 10.07 |
| 4 × 4 × 20.0 | Single detector | | 11.66 | 10.86 | 11.57 | 9.34 |

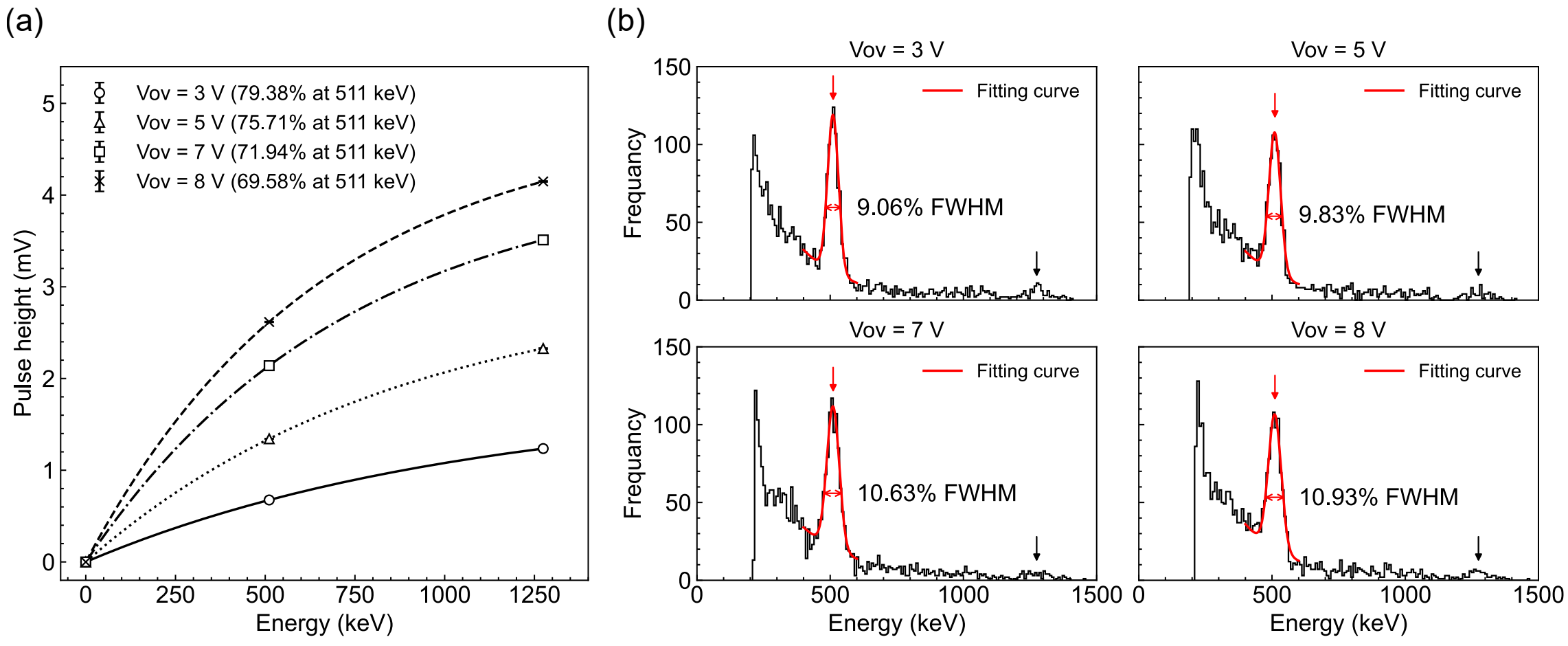

**Figure 6.** (a) Non-linearity curves and (b) energy spectra after non-linearity correction for channel

2 of the 4 mm$^2$ xDetector, which achieved the best performance at different overvoltages. The energy spectra display both the 511 keV (red arrow) and 1275 keV (black arrow) peaks from $^{22}$Na. The best linearity and energy resolution at 511 keV were 79.38% and 9.06% FWHM, respectively, at an overvoltage of 3 V.

### 4.3. Longitudinal spatial resolution

Table 3 presents the longitudinal spatial resolution of xDetectors with 3 and 4 mm$^2$ at overvoltages of 3 and 7 V, respectively. Each value represents the average resolution across all measurement positions. The center-of-gravity calculation using all channels provided better resolution compared to the ratio calculation between the two end channels. The best longitudinal spatial resolutions for the xDetectors with 3 and 4 mm$^2$ were 3.96 and 5.80 mm FWHM for channel 1 at an overvoltage of 3 V, respectively.

Figure 7(a) illustrates an example of eight signal waveforms from the xDetector with 3 mm$^2$ when a gamma ray interacts with a specific channel. The results indicate that channel 1, which has the highest pulse height, is the interaction channel, and the longitudinal position can be estimated at channel 6 based on the differences in pulse heights among channels 5 to 8. This observation aligns with the hypothesis described in Section 2, "*Concept of xDetector*." Figure 7(b) further shows the estimated longitudinal index at the scanned positions for the xDetector with 3 mm$^2$, which achieved the best performance.

**Table 3.** Longitudinal resolution of xDetectors using ratio and center-of-gravity calculations at overvoltages of 3 and 7 V. Each value represents the average resolution across all measurement positions.

| Crystal size (mm$^3$) | Composition | | Longitudinal spatial resolution (mm, FWHM) | | | |
|---|---|---|---|---|---|---|
| | | | Ratio calculation | | Center of gravity calculation | |
| | | | Vov = 3 (V) | Vov = 7 (V) | Vov = 3 (V) | Vov = 7 (V) |
| 3 × 3 × 12.8 | xDetector | Ch1 | 4.55 | 5.92 | **3.96** | 5.40 |
| | | Ch2 | 7.40 | 9.90 | 6.55 | 8.98 |
| | | Ch3 | 6.68 | 10.42 | 6.47 | 9.74 |
| | | Ch4 | 5.25 | 6.70 | 4.93 | 6.05 |
| 4 × 4 × 12.6 | xDetector | Ch1 | 5.57 | 6.32 | 5.80 | 6.47 |
| | | Ch2 | 7.46 | 8.39 | 7.25 | 8.18 |
| | | Ch3 | 6.01 | 6.86 | 5.77 | 6.82 |

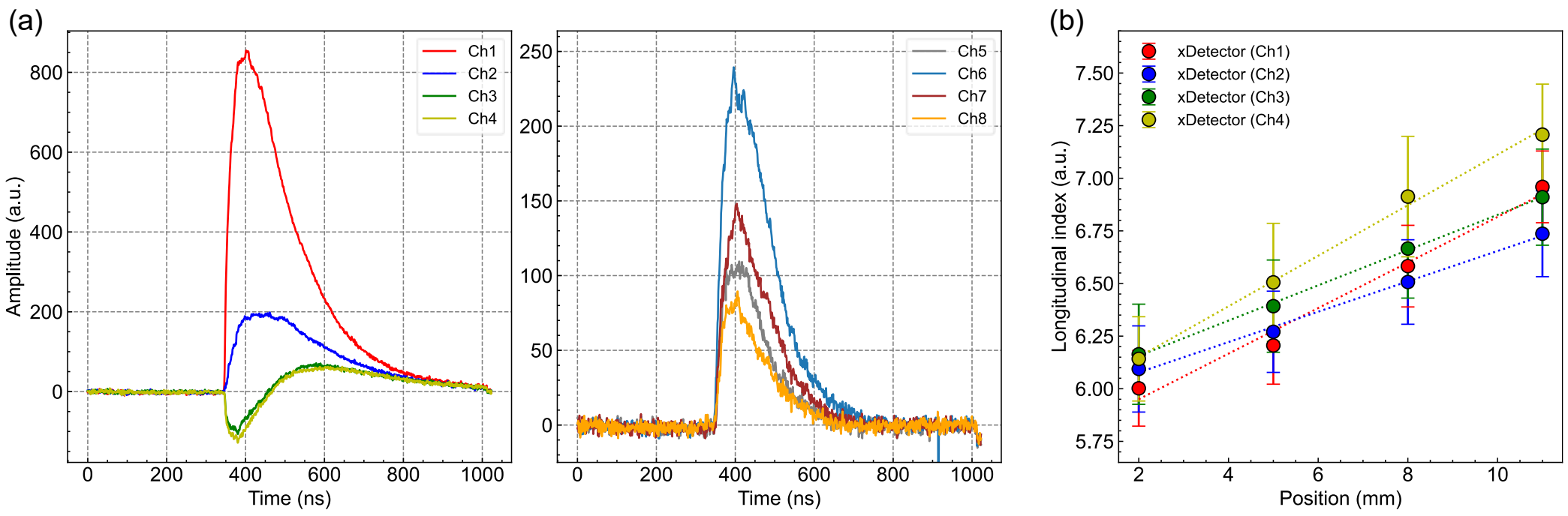

**Figure 7.** (a) Example of eight signal waveforms acquired from the xDetector with 3 mm$^2$ when a gamma ray interacts with a specific channel. Channel 1, with the highest pulse height, is identified as the interaction channel, and the longitudinal position is estimated at channel 6 based on the differences in pulse heights among channels 5 to 8. A portion of the waveforms appears negative due to a readout circuit issue. (b) Estimated longitudinal index at the scanned positions for the xDetector with 3 mm$^2$ that achieved the best performance. Error bars represent the standard deviation (sigma) of the fitted Gaussian distribution.

## 5. Discussion

In this study, we propose a readout scheme that orthogonally stacks two detector groups with one longitudinal side left uncovered by a reflector and facing the bare side to each other. The proposed xDetector achieves good timing resolution by reducing the scintillator thickness while maintaining sufficient sensitivity without the need for complex scintillator structures or SiPM couplings. Additionally, it can maintain spatial resolution by providing a scalable DOI capability. This approach suggests the potential for simultaneously addressing the multi-faceted trade-off between sensitivity, timing resolution, and spatial resolution that arises due to scintillator length. This is particularly important for the development of a high-resolution PET scanner targeted at specific organs within the human body (Akamatsu *et al* 2022, Onishi *et al* 2022, Li *et al* 2024) or small animals (Nagy *et al* 2013, Tomonari *et al* 2024).

As shown in figure 4, single detectors and xDetectors with a 12.6 and 12.8 mm length provided better CTR than the longer 20 mm detectors under all conditions due to the decreasing influence of PTTS. For the measurement of the single detector with dimensions 3 × 3 × 20 mm$^3$, the CTR was corrected due to the apparent discrepancy in LSO kinetics, while the CTR difference between the corrected 20 mm and the 12.8 mm detectors was larger than generally recognized. This discrepancy arises because the rise time and sigma in equation (6) were not fully considered in this study, and consequently, the calculation of the correction coefficient is uncertain to some extent. Although a simplified correction was performed because the single detector measurement for comparison was not the primary objective of this study, it is considered that the corresponding relationship of CTR would originally align with the detectors with 4 mm$^2$, as shown in figure 4(b).

The xDetector showed less degradation and equivalent CTR performance to conventional single detectors for the same crystal lengths, indicating that most photons are detected in an SiPM coupled to the interacting scintillator, with only a small amount of scintillation photons propagating to the other side of the detector group. This can be attributed to the use of air coupling between the detector groups.

The linearity of the detector is closely tied to the number of incident scintillation photons. Increasing the overvoltage and reducing the crystal length led to a deterioration in linearity, as shown in table 1 and figure 6(a). The linearity of the xDetectors was improved compared to conventional single detectors because some of the generated scintillation photons were propagated to upper- or lower-adjacent channels through air coupling, reducing the overall photon count. Conversely, the energy resolution of xDetectors tended to be slightly worse than that of conventional single detectors with the same length. However, the xDetector demonstrated comparable energy resolution to current PET scanners (Vandenberghe *et al* 2016).

The best longitudinal spatial resolution was achieved using the center-of-gravity calculation at an overvoltage of 3 V. Although the ratio-based calculation simplifies the readout electronics compared to the center-of-gravity calculation, its accuracy is limited because only two channels are used. As the calculation does not correct for non-linearity, better longitudinal resolution is obtained at lower applied voltages where linearity is relatively maintained. However, even with the readout scintillators stacked at 3 or 4 mm intervals in the longitudinal direction, the resolution remains worse than the size of the scintillators. Several factors may contribute to this degradation. The xDetector created in this study is held in a fixture, ensuring that the upper and lower scintillators are perfectly joined with no gaps between them. Ideally, an air gap would be created to deliberately induce total reflection of scintillation photons using differences in refractive index, but the current no-gap configuration might result in undesired smooth light leakage.

It is worth noting that the longitudinal information obtained from the xDetector has the potential to reduce biases in timing measurement due to uncertainties in photon propagation speed caused by differences in interaction positions within the scintillator (Shibuya *et al* 2008, Pizzichemi *et al* 2019, Toussaint *et al* 2019). Enhanced CTR with the longitudinal position correction would be indispensable when targeting CTR < 100 ps FWHM in the future.

The xDetector has two primary limitations. The first limitation is a reduced packing fraction, as the SiPMs of the xDetector are readout from the side of the detector module, as depicted in figures 1(d) and (e). Therefore, a reduced packing fraction can be expected depending on the thickness of the SiPM readout board. Assuming the thickness of an SiPM board is 0.2 mm, the packing fraction is calculated to be 96.9% compared to conventional 1-to-1 coupled detectors. Thus, developing a thin SiPM board will also be an area for future work. The second limitation is the increased number of SiPMs. The xDetector requires more SiPMs than the conventional 1-to-1 coupled detector. Although the exact number depends on both the readout pitch and the number of stacked layers, approximately 1.5 times more SiPMs will be required to maintain the same stopping power as conventional 20 mm-thick detectors. However, this number of required SiPMs is still significantly less than that needed in a dual-ended readout scheme.

This experiment demonstrated that the xDetector achieved a CTR of 175.3 ps FWHM, along with a linearity of 76.0% and an energy resolution of 11.1% FWHM at 511 keV, and a longitudinal spatial resolution of 3.96 mm FWHM, with eight 3 × 3 × 12.8 $mm^3$ LSO crystals coupled to MPPCs. Notably, there is still scope for performance improvement, as the xDetector was built using typical scintillator and SiPM configurations. The realization of the xDetector, combining state-of-the-art scintillators, photodetectors, and readout electronics, could potentially achieve a CTR of 100 ps FWHM without compromising sensitivity. The CTR correction performed in Section *4.1, "Coincidence Time Resolution (CTR)"* suggests a potential improvement in the CTR of the xDetector by 35% simply by replacing the LSO with an optimal one. Adequate selection of SiPMs in terms of higher PDE and faster SPTR also plays a crucial role in further improving CTR. Recent interdisciplinary work between SiPMs and nanophotonics can contribute to the development of future high-performance SiPMs (Mikheeva *et al* 2020, Enoch *et al* 2021, Uenoyama and Ota 2021, 2022). Finally, the use of a high frequency readout electronics (Cates *et al* 2018, Gundacker *et al* 2019, Jung *et al* 2024) can further enhance the CTR because we did not use any fast amplifiers in this study. These efforts could be also ultimately directed at CTR < 50 ps in the future. Such ultrafast timing enables reconstruction-free positron emission imaging (Kwon *et al* 2021, Onishi *et al* 2023, 2024) and will pave the way for a new perspective in the field of nuclear medicine.

## 6. Conclusion

In this study, we proposed a xDetector that features an orthogonally stacked readout scheme with one longitudinal side exposed, enabling the two detector groups to face each other directly. This design facilitates the maintenance of a thinner scintillator length while achieving better timing resolution compared to conventional detectors, without compromising sensitivity by stacking the detectors along the DOI axis of a PET scanner. Additionally, the proposed method provides scalable DOI capability, making it possible to develop a high-performance TOF-DOI PET detector. Experimental results demonstrated that the xDetector achieved a CTR of 175 ps FWHM, along with a linearity of 76.0% and an energy resolution of 11.1% FWHM at 511 keV, and a longitudinal resolution of 3.96 mm FWHM, using 3 × 3 × 12.8 $mm^3$ LSO crystals one-to-one coupled to MPPCs. The xDetector readout scheme, combined with state-of-the-art scintillators, photodetectors, and readout electronics, has the potential to achieve a CTR of 100 ps FWHM in a more practical configuration. Furthermore, the DOI information provides high spatial resolution, making it particularly effective for long axial field-of-view scanners.